\documentstyle[12pt]{article}
\font\titlefont=cmbx10 scaled \magstep4

\begin{document}

\begin{flushright}
\vspace*{-2cm}
gr-qc/9410043 \\ TUTP-94-16 \\ Oct. 21, 1994
\end{flushright}

\begin{center}
{\titlefont AVERAGED ENERGY} \\
\vspace {0.15 in}
{\titlefont CONDITIONS} \\
\vspace {0.15 in}
{\titlefont AND QUANTUM INEQUALITIES}\\
\vskip .3in
L.H. Ford\footnote{email: lford@pearl.tufts.edu} and
Thomas A. Roman\footnote{Permanent address: Department of Physics and Earth
Sciences, Central Connecticut State University, New Britain, CT 06050;
email: roman@ccsu.ctstateu.edu} \\
\vskip .2in
Institute of Cosmology\\
Department of Physics and Astronomy\\
Tufts University\\
Medford, Massachusetts 02155\\
\end{center}

\begin{abstract}
   In this paper, connections are uncovered between the averaged weak
(AWEC) and averaged null (ANEC) energy conditions, and quantum inequality
restrictions (uncertainty principle-type inequalities) on negative energy.
In two- and four-dimensional Minkowski spacetime, we examine quantized, free
massless, minimally-coupled scalar fields . In a
two-dimensional spatially compactified
Minkowski universe, we derive a covariant quantum inequality-type bound on the
difference of the expectation values of $T_{\mu\nu} u^{\mu} u^{\nu}$
in an arbitrary quantum state and in the Casimir vacuum state. From this
bound, it is shown that the difference of expectation values also obeys
AWEC and ANEC-type integral conditions. This is surprising, since it is
well-known that the expectation value of $T_{\mu\nu} u^{\mu} u^{\nu}$ in the
renormalized Casimir vacuum state alone satisfies neither quantum
inequalities nor averaged energy conditions. Such ``difference
inequalities'', if suggestive of the general case, might represent limits
on the degree of energy condition violation that is allowed over and above
any violation due to negative energy densities in a background vacuum state.
In our simple two-dimensional model, they provide physically interesting
examples of new constraints on negative energy which hold even when the
usual AWEC, ANEC, and quantum inequality restrictions fail. In the limit
when the size of the space is allowed to go to infinity, we derive quantum
inequalities for timelike and null geodesics which, in appropriate limits,
reduce to AWEC and ANEC in ordinary two-dimensional Minkowski spacetime.
Lastly, we also derive a covariant quantum inequality bound on the
energy density seen by an arbitrary inertial observer in four-dimensional
Minkowski spacetime. The bound implies that any inertial observer in flat
spacetime cannot see an arbitrarily large negative energy density which
lasts for an arbitrarily long period of time. From this latter bound, we
derive AWEC and ANEC.

\end{abstract}

\baselineskip=14pt

\section{Introduction}
\label{sec:intro}
       All known forms of classical matter obey the ``weak energy condition''
(WEC): $T_{\mu\nu} u^{\mu} u^{\nu} \geq 0$, for all timelike vectors $u^{\mu}$.
By continuity, this pointwise condition also holds for all null vectors.
Physically, this condition implies that the energy density of matter seen by
any observer is non-negative. The WEC is a crucial ingredient for ensuring
the focusing of null geodesic congruences in some
of the singularity theorems. Two notable examples are
Penrose's original 1965 \cite{P,HE} theorem, which predicts the
occurrence of a singularity at the endpoint of gravitational collapse, and
Hawking's extension of the theorem to open Friedmann-Robertson-Walker
universes \cite{H65}. More recently, the WEC has been used in proving
singularity
theorems for inflationary cosmologies and for certain classes of
closed universes \cite{BV1,BV2,B94}.

However, it was realized some time ago that
quantum field theory allows the violation of the WEC, as well as
all other known pointwise energy conditions \cite{EGJ}. Several examples of
situations involving negative energy densities and/or fluxes include: the
Casimir effect \cite{C,BM}, squeezed states of light \cite{WKHW},
radiation from moving mirrors \cite{FD}, the re-alignment of the
magnetic moments of an atomic spin system \cite{FGO},
and the Hawking evaporation of a black hole \cite{H75}. The experimental
observation of the first two effects indicates that we have to take the
idea of negative energy seriously. (Although, strictly
speaking, the energy density itself has not been measured.) On the other
hand, large violations of classical energy conditions might have drastic
consequences, such as a violation of the second law of thermodynamics
\cite{F78,D82} or of cosmic censorship \cite{FR90,FR92}.

Over the last several years, two approaches have emerged in the attempt
to constrain the extent of energy condition breakdown. The first, originally
introduced by Tipler \cite{T}, involves averaging the energy conditions over
timelike or null geodesics. An extension of some of Tipler's results shows
that Penrose's singularity theorem will still hold if the WEC is replaced
by an average of the WEC over certain half-complete null geodesics
\cite{TR}. For the purposes of this paper, and in keeping with the most
current usage, we will take the
``averaged weak energy condition'' (AWEC) to be the WEC
averaged over a complete timelike geodesic, i.e.,
\begin{equation}
\int_{-\infty}^{\infty}\,T_{\mu\nu} u^{\mu} u^{\nu}\, d\tau \geq 0\,.
                                                      \label{eq:AWEC}
\end{equation}
Here $u^{\mu}$ is the tangent vector to the timelike geodesic and $\tau$
is the observer's proper time.
Similarly the ``averaged null energy condition'' (ANEC) is taken to be the WEC
averaged over a complete null geodesic
\begin{equation}
\int_{-\infty}^{\infty}\,T_{\mu\nu} K^{\mu} K^{\nu}\, d\lambda \geq 0\,,
                                                      \label{eq:ANEC}
\end{equation}
where $K^{\mu}$ is the tangent vector to the null geodesic and $\lambda$
is an affine parameter. Borde has proven theorems on
the focusing of geodesics using other integral conditions than
those of Tipler \cite{B87}. His conditions only require that the relevant
integrals be periodically non-negative. A recent proof
of the positive mass theorem by Penrose, Sorkin, and Woolgar \cite {PSW} uses
Borde's integral focusing condition.
The theorems given in \cite{BV1,BV2,B94} can also be proved using
integral, as opposed to pointwise energy, conditions.

In addition to their importance in singularity theorems, averaged
energy conditions have recently become a topic of intense interest
because violation of ANEC has been shown to be a necessary requirement
for the maintenance of traversable wormholes \cite{MT}. Morris, Thorne,
and Yurtsever \cite{MTY} subsequently showed that if such wormholes can
exist, then one might be able to use them to construct time machines for
backward time-travel \cite{KT91/H92}. On the other hand, if ANEC is satisfied,
then the topological censorship theorem of Friedman, Schleich, and Witt
\cite{FSW} implies that one cannot actively probe multiply-connected
topologies. Any probe signal would get caught in the ``pinch-off''
of the topology.

The extent to which quantum field theory enforces averaged energy
conditions is not completely known. Klinkhammer has shown that ANEC is
satisfied when averaged along any complete null geodesic for quantized,
free scalar fields in four-dimensional Minkowski
spacetime \cite{K}. (This result is true for electromagnetic fields
as well \cite{Folacci,K2}.) He also showed that AWEC holds
when averaged along
any complete timelike geodesic, provided the coupling constant $\xi$ lies
within a certain range. This range includes the cases of minimal and
conformal coupling. However, Klinkhammer also observed that ANEC is
violated along every null geodesic in a two-dimensional spatially
compactified Minkowski spacetime for a quantized massless
scalar field in the Casimir vacuum state. A key feature of this
violation is the fact that, because of the periodic boundary conditions,
null geodesics in this spacetime are chronal, i.e., two points
on such a geodesic can be connected by a timelike curve. As a result, the
null geodesics can ``wrap around'' the space, and repeatedly traverse the
same negative energy region. By contrast, in a Casimir vacuum state with
vanishing (i.e., plate-type) boundary conditions, this cannot occur. However,
AWEC is violated in both two and four dimensions for
a static timelike observer in a Casimir vacuum
state, with either type of boundary conditions, since such an observer
simply sits in a region of constant negative energy density for all time.

Yurtsever has shown that ANEC holds on an arbitrarily curved
two-dimensional spacetime for a
conformally-coupled scalar field, provided the background spacetime satisfies
certain asymptotic regularity requirements \cite{Y}. Wald and Yurtsever
\cite{WY} have proven, among other things, that for a massless scalar
field in an arbitrary globally hyperbolic two-dimensional spacetime, ANEC
holds for all Hadamard states along any complete, achronal null geodesic.
They also show that, with a restriction on states, their results hold
for a massive scalar field in two-dimensional Minkowski spacetime and for
a massless or massive minimally-coupled scalar field in four-dimensional
Minkowski spacetime. However, they also show that ANEC cannot hold in
a general curved four-dimensional spacetime. In a recent elaboration of
the Wald-Yurtsever results, Visser \cite{Visser} has given a
sufficient condition for ANEC to be violated in a general spacetime.

The second approach toward determining the extent of energy condition
breakdown has taken the form of ``quantum inequality'' (QI)
restrictions, i.e., uncertainty-principle-type inequalities on the
magnitude and duration of negative energy fluxes due to quantum coherence
effects \cite{F78,F91}. For example, negative energy fluxes seen by
inertial observers in two-dimensional flat spacetime obey an inequality
of the form
\begin{equation}
|F|<\,(\Delta T)^{-2}\,,                      \label{eq:FT<1}
\end{equation}
where $|F|$ is the magnitude of the flux and $\Delta T$ is its duration. If
$|\Delta E|=|F|\, \Delta T$ is the amount of negative energy passing by a
fixed spatial position in a time $\Delta T$, then Eq.~(\ref{eq:FT<1})
implies that
\begin{equation}
|\Delta E|\, \Delta T<1 \,.
\end{equation}
Therefore, $|\Delta E|$ is less than the quantum uncertainty in the energy,
$(\Delta T)^{-1}$, on the timescale $\Delta T$.

     In Ref. \cite{F91} a more
rigorous form of this kind of inequality was proven to hold for all quantum
states of a free minimally-coupled massless scalar field in both two
and four-dimensional Minkowski spacetime. The more precise inequalities
are expressed as an integral of the energy flux multiplied
by a ``sampling function'', i.e., a peaked function of time
whose time integral is unity and whose characteristic width is $t_0$. A
suitable
choice of such a function is $t_0/[\pi(t^2+{t_0}^2)]$. If we define
the integrated flux, $\hat F$, by
\begin{equation}
{\hat F} \equiv {t_0 \over \pi} \int_{-\infty}^{\infty}
{{F(t) dt}\over {t^2+{t_0}^2}}\,,               \label{eq:DEF/HATF}
\end{equation}
then these inequalities can be written as
\begin{equation}
{\hat F} \geq -{1 \over {16 \pi {t_0}^2}},
                                                \label{eq:HATF/2D}
\end{equation}
and
\begin{equation}
{\hat F} \geq -{3 \over {32 {\pi}^2 {t_0}^4}}\,,
                                                \label{eq:HATF/4D}
\end{equation}
for all $t_0$, in two- and four-dimensions, respectively. These
inequalities are of the
form required to prevent large-scale violations of the second law of
thermodynamics. Similar inequalities were found to hold for a quantized
massless, minimally-coupled scalar field propagating on two- and
four-dimensional extreme Reissner-Nordstr{\o}m black hole backgrounds. These
inequalities were shown to be sufficient to foil attempts at
creating an unambiguous violation of cosmic censorship by injecting a negative
energy flux into an extreme charged black hole \cite{FR90,FR92}.
It is important to note that
{\it the energy-time uncertainty principle was not used to derive any of
these QI restrictions}. They arise directly from quantum
field theory.

There has been as yet no concerted effort to link these two different
approaches to constraining energy condition violations. Perhaps this is
due in part to the fact that, in contrast to the averaged
energy conditions, the QI's
Eq.~(\ref{eq:HATF/2D}) and Eq.~(\ref{eq:HATF/4D}) are not written in a
covariant form. Yet in some cases,
the QI's yield stronger restrictions than the averaged
energy conditions. Consider the following example \cite{FR92}. In a
flat two-dimensional spacetime, an inertial observer encounters a
$\delta$-function pulse of negative $(-)$ energy, with magnitude $|\Delta E|$,
followed a time $T$
later by a compensating similar pulse of positive $(+)$ energy. The AWEC simply
requires that the compensating $(+)$ energy must arrive at {\it some}
time after the incidence of the $(-)$ energy, perhaps arbitrarily far
in the future. On the other hand the QI,
Eq.~(\ref{eq:HATF/2D}), implies \cite{F91} that the positive energy must
arrive {\it no later than} a time $T<{(|\Delta E|)}^{-1}$. A hint that
links between QI's and averaged energy conditions
might exist can be found in Eq. (59) of \cite{WY}.

In the present paper, we show that there is in fact a deep connection
between AWEC, ANEC, and QI-
type restrictions on negative energy. We show that, in a number of cases, it is
possible to derive AWEC and ANEC from QI's on timelike
and null geodesics. The paper is organized as follows.
In Sec.~\ref{sec:2Ddiff}, we consider
a quantized, massless scalar field in a 2D spatially compactified
Minkowski spacetime with circumference $L$. We derive a QI
on the {\it difference} between
the expectation values of $T_{\mu\nu} u^{\mu} u^{\nu}$ in an arbitrary
quantum state and in the Casimir vacuum state. Here $u^{\mu}$ is the
two-velocity of an arbitrary inertial observer. In the $L\rightarrow\infty$
limit, we obtain a covariant QI for the energy density
averaged over an arbitrary timelike geodesic in 2D
(un-compactified) Minkowski spacetime. By then letting the width of the
sampling function go to infinity, we derive AWEC in 2D Minkowski spacetime.
Surprisingly, we find that the difference of expectation values, in the
{\it compactified} Minkowski spacetime, also obeys an AWEC-type integral
inequality. This is interesting because AWEC is violated for
$\langle T_{\mu\nu} u^{\mu} u^{\nu} \rangle$ in the (renormalized) Casimir
vacuum state by itself. Such ``difference inequalities''
might provide new measures of the degree of energy condition
violation in cases where the usual averaged energy conditions and QI's
fail \cite{FR93}. In Sec.~\ref{sec:2DNULL}, we derive analogous
QI's for null geodesics. We also show how the inequality for null geodesics
in 2D uncompactified Minkowski spacetime can be obtained
from the original flux inequality, Eq.~(\ref{eq:HATF/2D}).
{}From these QI's for null geodesics, we derive ANEC. Again we
find that in the finite $L$ case, an ANEC-type integral inequality holds for
the difference of expectation values. In Sec.~\ref{sec:4DQI},
we derive a covariant QI on
the energy density of a quantized, free massless, minimally-coupled scalar
field in 4D (uncompactified) Minkowski spacetime. This inequality,
originally conjectured in \cite{F91}, is analogous to the one for
energy flux in 4D, Eq.~(\ref{eq:HATF/4D}). We also show that AWEC and
ANEC can be obtained in suitable limits of our inequality. A summary
of our results is contained in Sec.~\ref{sec:summary}.
Units with $\hbar=c=1$ will be used.

\section{A ``Difference Inequality'' in 2D}
\label{sec:2Ddiff}
   Consider the difference of expectation values defined by:
\begin{equation}
D\langle T_{\mu\nu}u^\mu u^\nu\rangle \equiv
 {\langle \psi\mid T_{\mu\nu}u^\mu u^\nu \mid\psi\rangle-
 \langle 0_{C}\mid T_{\mu\nu}u^\mu u^\nu \mid 0_{C}\rangle}\,,  \label{eq:Ddef}
\end{equation}
where $\mid\psi\rangle$ is an arbitrary quantum state, $\mid 0_{C}\rangle$
is the Casimir vacuum state, and $u^\mu$ is the unit tangent vector to an
arbitrary timelike geodesic. The effects of negative energy can be
adjusted in two ways: a) by changing the quantum state $\mid\psi\rangle$,
and/or b) by changing the energy density in the background
Casimir vacuum state.
The latter could be accomplished by changing the plate separation in the case
of vanishing (i.e., plate-type) boundary conditions, or in the case of
periodic boundary conditions by altering the size of the compactified space.
The difference of expectation values, Eq.~(\ref{eq:Ddef}), may be regarded
as a measure of the negative energy density over and above that of the
background Casimir vacuum energy density.

    Let us take the stress-energy tensor operator to be that of a massless
scalar field in a two-dimensional cylindrical Minkowski spacetime:
\begin{equation}
T_{\mu \nu} = \phi_{,\mu}\phi_{,\nu}
-{1\over 2}\eta_{\mu \nu}\phi_{,\alpha}\phi^{,\alpha}\, .  \label{eq:ST}
\end{equation}
The field operator may be expanded in terms of creation and annihilation
operators as
\begin{equation}
\phi = {\sum_k} ({a_k}{f_k} + {a_k}^\dagger{f_k}^\ast). \label{eq:phi}
\end{equation}
Here the mode functions are taken to be
\begin{equation}
f_k = {i \over \sqrt{2\omega L}} e^{i(kx - \omega t)}, \label{eq:fk}
\end{equation}
where $\omega = |k|$ and periodicity of length $L$ has been imposed in the
spatial direction, so that $k$ takes on the discrete values
\begin{equation}
k={{2\pi l}\over L},\;\;l=\pm1,\pm2\ldots\, .      \label{eq:kmodes}
\end{equation}

    The expansion of the field operator described by Eqs.~(\ref{eq:phi})
and ~(\ref{eq:fk}) is not complete in that it omits the zero mode which
exists for the massless scalar field on a compactified spacetime.
However, as is discussed in Ref. \cite{FP}, this mode always gives a positive
contribution to the energy density. (The magnitude of this contribution
depends upon an additional parameter which must be specified in order to
uniquely define the ground state in the compactified space.) Here our
aim is to derive a lower bound involving the difference defined in
Eq.~(\ref{eq:Ddef}), and we will ignore the zero mode contribution.
Inclusion of the zero mode can only increase this difference, hence
the lower bounds so obtained will always be satisfied.

     The closure of the spatial dimension introduces a preferred reference
frame.  Let us now
consider an inertial observer who moves with velocity $V$ relative to
this frame so that $u^\mu = \gamma(1,V)$, where
$\gamma=(1-V^2)^{-1/2}$. Therefore,
\begin{equation}
T_{\mu\nu}u^\mu u^\nu={1\over 2}\left({{1+V^2}\over{1-V^2}}\right)\,
[(\phi,_{t})^2 +(\phi,_{x})^2]\,+\left({V\over{1-V^2}}\right)\,
[\phi,_{t}\phi,_{x}+\phi,_{x}\phi,_{t}].
                                            \label{eq:STOP}
\end{equation}
The observer's worldline is given by
\begin{equation}
x=(x_0+Vt) \mp mL,\qquad t=\tau (1-V^2)^{-1/2},    \label{eq:OBWL}
\end{equation}
where $\tau$ is the observer's proper time, and $m$ is an integer (i.e.,
the winding number). The $(-)$ sign applies if $V>0$ and the $(+)$ sign if
$V<0$. For simplicity, we will set $x_0=0$. Combining the
previous equations yields
\begin{eqnarray}
T_{\mu\nu} {u^\mu} {u^\nu}&=&{1\over {2L}} {\sum_{k,k'}}
{1\over{\sqrt{\omega \omega'}}}\,
\left[{1\over 2}\left({{1+V^2}\over{1-V^2}}\right)(\omega' \omega + k k')
-\left({V\over{1-V^2}}\right)(\omega' k + \omega k')\right] \nonumber \\
                           &\times& \left[ ({a_{k'}}{a_{k}})
{e^{i\{(k'+k)[V\tau(1-V^2)^{-1/2} \mp mL]-
(\omega' + \omega)\tau(1-V^2)^{-1/2}\}}}\right.    \nonumber \\
&+& ({{a_{k'}}^\dagger}{a_{k}})
{e^{-i\{(k'-k) [V\tau(1-V^2)^{-1/2} \mp mL] -
(\omega' - \omega)\tau(1-V^2)^{-1/2}\}}}    \nonumber \\
&+& \left. h.c.'s +\delta_{k'k} \right],      \label{eq:NEXTSTOP}
\end{eqnarray}
where the $h.c.'s$ are hermitian conjugates.

If we now split the modes into right-moving ($k>0$) and left-moving ($k<0$),
then there will be four possible combinations, corresponding to the four
sums in Eq.~(\ref{eq:NEXTSTOP}). These are: $(k'>0,k>0);(k'<0,k<0);(k'>0,
\linebreak k<0);(k'<0,k>0)$.
In all cases $\omega'=|k'|,\omega=|k|$. From the form of the first term in
square brackets in Eq.~(\ref{eq:NEXTSTOP}), it is easy to see that the only
non-trivial combinations are the two where $k',k$ have
the same sign. We can also see
that the $m$-dependence drops out. This is because it always appears in the
form of exponentials, such as $e^{ i(k' + k)mL}$, which all equal $1$,
since from Eq.~(\ref{eq:kmodes}), $(l' \pm l)m$ is an integer. Therefore,
Eq.~(\ref{eq:NEXTSTOP}) becomes
\begin{eqnarray}
T_{\mu\nu} {u^\mu} {u^\nu}&=& {1\over
{2L}}\left\{\left({{1-V}\over{1+V}}\right)
{\sum_{k',k>0}}\sqrt{k'k}
\left[ ({a_{k'}}{a_{k}}){e^{-i(k'+k)\sqrt{{{1-V}\over{1+V}}}\tau}}
\right.\right. \nonumber \\
&+& \left.({{a_{k'}}^\dagger}{a_{k}})
{e^{i(k'-k)\sqrt{{{1-V}\over{1+V}}}\tau}}+ h.c.'s
+\delta_{k'k} \right] \nonumber \\
&+& \left({{1+V}\over{1-V}}\right)
{\sum_{k',k<0}}\sqrt{k'k}
\left[ ({a_{k'}}{a_{k}}){e^{i(k'+k)\sqrt{{{1+V}\over{1-V}}}\tau}}\right.
\nonumber \\
&+& \left.\left.({{a_{k'}}^\dagger}{a_{k}})
{e^{-i(k'-k)\sqrt{{{1+V}\over{1-V}}}\tau}}
+ h.c.'s +\delta_{k'k} \right] \right\}.                 \label{eq:LASTSTOP}
\end{eqnarray}

  Our goal is to formulate and prove an inequality involving a (proper) time
integral of $D\langle T_{\mu\nu}u^\mu u^\nu\rangle$. The expectation value
in the Casimir vacuum,
$\langle 0_{C}\mid T_{\mu\nu}u^\mu u^\nu \mid 0_{C}\rangle$, is given by the
$\delta_{k'k}$ terms of Eq.~(\ref{eq:LASTSTOP}). Thus
$D\langle T_{\mu\nu}u^\mu u^\nu\rangle$ is the expectation value of
Eq.~(\ref{eq:LASTSTOP}) omitting these terms. Following Ref. \cite{F91},
we multiply $D\langle T_{\mu\nu}u^\mu u^\nu\rangle$ by a peaked function of
proper time whose time integral is unity and whose characteristic width
is $\tau_0$. Such a function is ${\tau}_0 /[\pi({\tau}^2+{{\tau}_0}^2)]$.
Define
$\hat D\langle T_{\mu\nu}u^\mu u^\nu\rangle$, by
\begin{eqnarray}
&&{\hat D\langle T_{\mu\nu}u^\mu u^\nu\rangle}
\equiv {\tau_0 \over \pi} \int_{-\infty}^{\infty}
{{D\langle T_{\mu\nu}u^\mu u^\nu\rangle d{\tau}}\over {{\tau}^2+{{\tau}_0}^2}}
 =      \nonumber \\
&&\!\!\!\!\!\!\!\!\!\!\!\!
{1\over {L}}\left\{ \left({{1-V}\over{1+V}}\right)
Re{\sum_{k',k>0}}\sqrt{k'k}
\left[ \langle{a_{k'}}{a_{k}}\rangle
{e^{-(k'+k)\sqrt{{{1-V}\over{1+V}}}\tau_0}}
+ \langle{{a_{k'}}^\dagger}{a_{k}}\rangle
{e^{-|k'-k|\sqrt{{{1-V}\over{1+V}}}\tau_0}}\right] \right. +     \nonumber \\
&&\!\!\!\!\!\!\!\!\!\!\!\! \left({{1+V}\over{1-V}}\right)
Re{\sum_{k',k<0}}\sqrt{k'k}
\left[ \langle{a_{k'}}{a_{k}}\rangle
{e^{-|k'+ k|\sqrt{{{1+V}\over{1-V}}}\tau_0}}
+ \left.\langle{{a_{k'}}^\dagger}{a_{k}}\rangle
{e^{-|k'-k|\sqrt{{{1+V}\over{1-V}}}\tau_0}}\right]
\right\}\,.                                           \label{eq:DHAT}
\end{eqnarray}

  For the moment, let $T_0\equiv \sqrt{{{1-V}\over{1+V}}}\tau_0$ and
$\bar T_0\equiv \sqrt{{{1+V}\over{1-V}}}\tau_0$. Then the first sum can be
written as
\begin{eqnarray}
\hat S_1 &\equiv&
{1\over {L}}\left({{1-V}\over{1+V}}\right)
Re{\sum_{k',k>0}}\sqrt{k'k} \left[ \langle{a_{k'}}{a_{k}}\rangle
{e^{-(k'+k)T_0}} + \langle{{a_{k'}}^\dagger}{a_{k}}
\rangle {e^{-|k'-k|T_0}}\right] \nonumber \\
&\geq& {1\over {L}}\left({{1-V}\over{1+V}}\right)
Re{\sum_{k',k>0}}\sqrt{k'k} \, {e^{-(k'+k)T_0}}\,
\left[ \langle{a_{k'}}{a_{k}}\rangle
 + \langle{{a_{k'}}^\dagger}{a_{k}}\rangle \right],         \label{eq:S1}
\end{eqnarray}
where we have used the lemma in Appendix B of \cite{F91}. Now let $h(k)=
\sqrt k\,{e^{-kT_0}}$. Then by applying the lemma in Appendix A of \cite{F91}
to the right hand side of Eq.~(\ref{eq:S1}), we have
\begin{equation}
\hat S_1 \geq -{1\over {2L}}\left({{1-V}\over{1+V}}\right) {\sum_{k>0}}\,
{h^2}(k)=-{1\over {2L}}\left({{1-V}\over{1+V}}\right) {\sum_{k>0}}\,
k{e^{-2kT_0}}.
                                                         \label{eq:S1INQ}
\end{equation}
The fact that
\begin{equation}
-{1\over {2L}} {\sum_{k>0}}\, k{e^{-2kT_0}}
=-{\pi \over 2 L^2}\, {1 \over {\left[cosh\left({{4\pi}\over L}
{\sqrt{{{1-V}\over{1+V}}}\tau_0}\right)-1\right]}}\: ,
\label{eq:KSUMCOSH}
\end{equation}
implies
\begin{equation}
\hat S_1 \geq
-{\pi \over {2 L^2}}\,\left({{1-V}\over{1+V}}\right) \,
{1 \over {\left[cosh\left({{4\pi}\over L}
{\sqrt{{{1-V}\over{1+V}}}\tau_0}\right)-1\right]}} \:.   \label{eq:LASTS1INQ}
\end{equation}
We obtain a similar inequality on the $(k',k<0)$ sum, $\hat S_2$, only
with $V \rightarrow -V$. If we combine the inequalities on $\hat S_1$ and
$\hat S_2$, we finally get our desired result
\begin{eqnarray}
{\hat D\langle T_{\mu\nu}u^\mu u^\nu\rangle}
&\equiv& {\tau_0 \over \pi} \int_{-\infty}^{\infty}
{{\left[{\langle \psi\mid T_{\mu\nu}u^\mu u^\nu \mid\psi\rangle-
 \langle 0_{C}\mid T_{\mu\nu}u^\mu u^\nu \mid 0_{C}\rangle}\right]
d{\tau}}\over
{{\tau}^2+{{\tau}_0}^2}} \nonumber \\
&\geq& -{\pi \over {2 L^2}}\,\left\{ \left({{1-V}\over{1+V}}\right) \,
{1 \over {\left[cosh\left({{4\pi}\over L}
{\sqrt{{{1-V}\over{1+V}}}\tau_0}\right)-1\right]}}\right. \nonumber \\
&+& \left.\left({{1+V}\over{1-V}}\right) \,
{1 \over {\left[cosh\left({{4\pi}\over L}
{\sqrt{{{1+V}\over{1-V}}}\tau_0}\right)-1\right]}} \right\}.
\label{eq:DHATINQ}
\end{eqnarray}
This inequality holds for all choices of $\tau_0$.

There are a number of interesting limits of Eq.~(\ref{eq:DHATINQ}), which we
now discuss. For an observer who is static relative to the preferred frame,
$V\rightarrow 0$, and our expression reduces to
\begin{equation}
{\hat D\langle T_{\mu\nu}u^\mu u^\nu\rangle} \geq
-{\pi \over L^2}\, {1 \over {\left[cosh\left({{4\pi}\over L}
{\tau_0}\right)-1\right]}}.                          \label{eq:DHATSTAT}
\end{equation}

In the limit as $L\rightarrow\infty$, for fixed $\tau_0$ and $V$,
from the fact that $cosh(x)\simeq 1+{1\over 2}x^2+\dots$, for $|x|\ll 1$,
the $V$-dependence cancels out and we have
\begin{equation}
{\hat D\langle T_{\mu\nu}u^\mu u^\nu\rangle} \geq
-{1\over {8\pi {\tau_0}^2}}.                           \label{eq:DHAT/L}
\end{equation}
However, in the $L\rightarrow \infty$ limit,
${\langle \psi\mid T_{\mu\nu}u^\mu u^\nu \mid\psi\rangle-
 \langle 0_{C}\mid T_{\mu\nu}u^\mu u^\nu \mid 0_{C}\rangle}$ simply reduces
to ${\langle T_{\mu\nu}u^\mu u^\nu \rangle}$, where the
expectation value is taken in the state\linebreak $\mid\psi\rangle$ and
the operator $T_{\mu\nu}u^\mu u^\nu$ is normal-ordered with respect to the
usual Minkowski vacuum state. Equation~(\ref{eq:DHAT/L}) then becomes
\begin{equation}
{\tau_0 \over \pi} \int_{-\infty}^{\infty}
{{\langle T_{\mu\nu}u^\mu u^\nu\rangle d{\tau}}\over {{\tau}^2+{{\tau}_0}^2}}
\geq -{1\over {8\pi {{\tau_0}^2}}},                    \label{eq:TLQI2DUNCP}
\end{equation}
for all $\tau_0$. This quantum inequality restriction on the energy density
in two-dimensional uncompactified Minkowski space is analogous to a similar
inequality, Eq.~(\ref{eq:HATF/2D}), derived in \cite{F91},
on the energy flux measured by a timelike inertial observer in 2D.
If we now take the limit of Eq.~(\ref{eq:TLQI2DUNCP}) as
${\tau_0}\rightarrow \infty$, we derive AWEC, Eq.~(\ref{eq:AWEC}),
in ordinary two-dimensional Minkowski spacetime.

If we take the limit of Eq.~(\ref{eq:DHATINQ}) when
${\tau_0}\rightarrow \infty$, for fixed $L$ and $V$, we obtain
\begin{eqnarray}
&&\int_{-\infty}^{\infty} D\langle T_{\mu\nu}u^\mu u^\nu\rangle d{\tau}=
                                                       \nonumber \\
&&\int_{-\infty}^{\infty}
{\left[\langle \psi\mid T_{\mu\nu}u^\mu u^\nu \mid\psi
\rangle- \langle 0_{C}\mid T_{\mu\nu}u^\mu u^\nu \mid 0_{C}\rangle\right]}
d{\tau} \geq 0\,,                                        \label{eq:AWECL2D}
\end{eqnarray}
i.e., an AWEC-type integral holds for the {\it difference} of the expectation
values. Note that Eq.~(\ref{eq:AWECL2D}) also holds in the $L \rightarrow 0$
limit of Eq.~(\ref{eq:DHATINQ})
for fixed $\tau_0$ and $V$. By contrast, an AWEC integral on
${\langle 0_{C}\mid T_{\mu\nu}u^\mu u^\nu \mid 0_{C}\rangle}_{ren}$, the
renormalized vacuum expectation value of a massless scalar field in the
Casimir vacuum state, is {\it not} satisfied \cite{K}.

Lastly, in the limit ${\tau_0}\rightarrow 0$, for fixed $L,V$, we
get the rather weak bound that
\begin{equation}
{\hat D\langle T_{\mu\nu}u^\mu u^\nu\rangle} \geq -{\infty}.
                                                       \label{eq:DHAT/TAU/0}
\end{equation}
The limit of $V\rightarrow\pm 1$ will be discussed in the next section.

\section{A Derivation of a QI for Null Geodesics in 2D}
\label{sec:2DNULL}

    In this section, we will derive a QI for null
geodesics in both 2D compactified and ordinary 2D Minkowski spacetime using two
different methods. As a limiting case, we obtain ANEC.
  Let us repeat the analysis of the last section for null geodesics. We take
the tangent vector to our null geodesic to be given by $K^{\mu}=
\alpha\,(1,\pm 1)$, where the plus sign is taken for a right-moving geodesic
and the minus sign for one which is left-moving, and where $\alpha$ is an
arbitrary positive constant. A null geodesic in 2D compactified
Minkowski spacetime is given by:
\begin{equation}
x= x_0 \pm (t-mL)\,,                       \label{eq:NULLX}
\end{equation}
for right-moving and left-moving geodesics, respectively, where $m$ again
is the winding number. For convenience, we choose $x_0=0$. From the null
geodesic equation we have that ${dt}/{d\lambda}= \alpha$, so
\begin{equation}
t= \alpha \lambda\, + c_0\,.                 \label{eq:NULLAFF}
\end{equation}
where $c_0$ is a constant which, for simplicity, we choose to be $0$ so that
$\lambda=0$ when $t=0$.

   We now take our geodesic to be left-moving, i.e., we choose the minus sign
in Eq.~(\ref{eq:NULLX}). The analog of Eq.~(\ref{eq:LASTSTOP}) is
\newpage
\begin{eqnarray}
&&\!\!\!\!\!\!\!\!\!\!\!\!\!\!\!\! T_{\mu\nu} {K^\mu} {K^\nu} = \nonumber \\
&&\!\!\!\!\!\!\!\!\!\!\!\!\!\!\!\!
{ {2{\alpha}^2}\over {L} } \, {\sum_{k',k>0}}\sqrt{k'k}
\left[ ({a_{k'}}{a_{k}}){e^{-2i \alpha \lambda (k'+k)} }
+ ({{a_{k'}}^\dagger}{a_{k}})
{e^{2i \alpha \lambda (k'-k)} }+ h.c.'s
+\delta_{k'k} \right]\,.
                                                 \label{eq:TKK}
\end{eqnarray}
Note that in contrast to Eq.~(\ref{eq:LASTSTOP}), only the right-moving modes
contribute, i.e., those which move in the direction opposite to that
of the null geodesic. A repetition of the method used to obtain
Eq.~(\ref{eq:DHATINQ}) yields
\begin{equation}
{\hat D\langle T_{\mu\nu} K^\mu K^\nu\rangle} \, \equiv \,
{\lambda_0 \over \pi} \int_{-\infty}^{\infty}
{ D {\langle T_{\mu\nu}K^\mu K^\nu\rangle d{\lambda}}\over
{{\lambda}^2+{{\lambda}_0}^2}}   \geq
-{ {2\pi {\alpha}^2} \over L^2}\,
{1 \over
{\left[cosh\left({{8\pi \over L} \alpha {\lambda}_0} \right)-1\right]}}\, ,
                                               \label{eq:DHATTKK}
\end{equation}
for all $\lambda_0$.
Let ${\lambda}_p= L/ {\alpha}$, which is the affine parameter separation
between points on the null geodesic which are at the same spatial location.
We may rewrite Eq.~(\ref{eq:DHATTKK}) as
\begin{equation}
{\hat D\langle T_{\mu\nu} K^\mu K^\nu\rangle}  \geq
-{ {2\pi} \over {{\lambda}_p}^2}\,
{1 \over
{\left[cosh
\left( 8\pi { {{\lambda}_0}/ {{\lambda}_p} } \right)-1\right]}}\, ,
                                               \label{eq:DHATTKK2}
\end{equation}
for all $\lambda_0$.
Note that Eq.~(\ref{eq:DHATTKK2}) is invariant under rescaling of the affine
parameter. That is, if $\lambda$, $\lambda_0$, and $\lambda_p$ are multiplied
by the same constant, the form of the equation is unchanged \cite{CSCALE}.

  If we now take the $L \rightarrow \infty$ limit of Eq.~(\ref{eq:DHATTKK}),
we obtain
\begin{equation}
{\lambda_0 \over \pi} \int_{-\infty}^{\infty}
{{\langle T_{\mu\nu}K^\mu K^\nu\rangle d{\lambda}}\over
{{\lambda}^2+{{\lambda}_0}^2}}   \geq -{1\over {16\pi {{\lambda_0}^2}}}\, ,
                                                  \label{eq:NQI1}
\end{equation}
for all $\lambda_0$.
This inequality is the null analog of Eq.~(\ref{eq:TLQI2DUNCP}). The factor
of $2$ difference on the right-hand sides is due to the fact that only
right-moving modes contribute in the case of Eq.~(\ref{eq:NQI1}).
If we now take the limit of Eq.~(\ref{eq:NQI1}) as
${\lambda_0}\rightarrow\infty$, which corresponds to sampling the entire
null geodesic, we get ANEC:
\begin{equation}
\int_{-\infty}^{\infty}
{\langle T_{\mu\nu}K^\mu K^\nu\rangle d{\lambda}} \geq 0.  \label{eq:ANEC2D}
\end{equation}

  If we take the limit of Eq.~(\ref{eq:DHATTKK}) as
${\lambda_0}\rightarrow \infty$, for fixed $L$, we obtain the null analog of
Eq.~(\ref{eq:AWECL2D}):
\begin{eqnarray}
&&\int_{-\infty}^{\infty} D\langle T_{\mu\nu}K^\mu K^\nu\rangle d{\lambda} =
                                                       \nonumber \\
&&\int_{-\infty}^{\infty}{\left[\langle \psi\mid T_{\mu\nu}K^\mu K^\nu \mid\psi
\rangle- \langle 0_{C}\mid T_{\mu\nu}K^\mu K^\nu \mid 0_{C}\rangle\right]}
d{\lambda} \geq 0\,.                                        \label{eq:ANECL2D}
\end{eqnarray}
Thus we see that an ANEC-type inequality holds for the {\it difference} of
the expectation values, even though ANEC for
${\langle 0_{C}\mid T_{\mu\nu}K^\mu K^\nu \mid 0_{C}\rangle}_{ren}$, by
itself, is {\it not} satisfied. Remarkably, this difference inequality
appears to hold even though the null geodesics in this spacetime are chronal.
The chronality of null geodesics in 2D compactified Minkowski spacetime was
a key feature of Klinkhammer's observation \cite{K} that ANEC is violated
in this spacetime for
${\langle 0_{C}\mid T_{\mu\nu}K^\mu K^\nu \mid 0_{C}\rangle}_{ren}$.
 It is of interest to note that QI's
for timelike and null geodesics hold for the difference of the
expectation values, although they are not satisfied for
${\langle 0_{C}\mid T_{\mu\nu}\mid 0_{C}\rangle}_{ren}$ alone.

    We now wish to show how Eq.~(\ref{eq:DHATTKK}) may be derived by taking
the null limit of a difference inequality for timelike observers. From our
previous discussion, we know that in the null case only modes moving
in the opposite direction to that of the geodesic (which we will again take
to be to the left) will contribute to the bound.
Thus let us start with the analog of Eq.~(\ref{eq:DHATINQ}) which applies
for a quantum state in which only modes moving to the right are excited:
\begin{equation}
{\hat D\langle T_{\mu\nu}u^\mu u^\nu\rangle} \geq
 -{\pi \over {2 L^2}}\,
\left({{1-V}\over{1+V}}\right) \,
{1 \over {\left[cosh\left({{4\pi}\over L}
{\sqrt{{{1-V}\over{1+V}}}\tau_0}\right)-1\right]}} \,.  \label{eq:DHATRIGHT}
\end{equation}
Let
\begin{equation}
\lambda=\gamma\tau/ \alpha,\qquad {\lambda_0}=\gamma{\tau_0}/ \alpha \,,
                                                   \label{eq:TAULAMB}
\end{equation}
where $\alpha$ is the arbitrary constant corresponding to our scaling of the
affine parameter. We first rewrite Eq.~(\ref{eq:DHATRIGHT}) using
Eq.~(\ref{eq:TAULAMB}) to replace $\tau$ and $\tau_0$ by $\lambda$ and
$\lambda_0$, respectively. Divide both sides by $\gamma^2/\alpha^2$
and take the null limit of this expression,
i.e., one in which ${\gamma=(1-V^2)^{-1/2}}\rightarrow\infty$ and
$\tau\rightarrow 0$ as $V\rightarrow -1$, such that the product $\gamma\tau$
remains finite \cite{NULLLIMITCOMMENT}. In this limit, $\lambda$ becomes an
affine parameter for our null geodesic.
(Note that this is {\it not} a fixed $\tau_0$ limit.)
If we rewrite the left-hand side in terms of $K^{\mu}$, the result is
Eq.~(\ref{eq:DHATTKK}).

Alternatively, we could have derived the inequality for uncompactified
spacetime, Eq.~(\ref{eq:NQI1}), from the 2D
inequality on energy fluxes, Eq.~(\ref{eq:HATF/2D}), given in \cite{F91}.
Let $u^\mu = \gamma(1,V)$
be the two-velocity of an inertial observer. The flux in
this observer's frame is given by
\begin{equation}
F = - T_{\mu\nu} {u^\mu} {n^\nu}.    \label{eq:flux}
\end{equation}
Here $n^\mu= \gamma(V,1)$ is a spacelike unit normal
vector, for which ${n_\mu}{u^\mu} = 0$.
Consider the most general quantum state in which only
particles moving in the $+x$ direction are present. A negative energy flux then
arises if the energy flow is in the $-x$ direction. Examine the energy
flux at an arbitrary spatial point, which we take to be $x = 0$,
for an observer with $u^\mu=(1,0)$. In Ref. \cite{F91} the inequality,
Eq.~(\ref{eq:HATF/2D}), was shown to be satisfied for this observer.
However, because of the underlying Lorentz-invariance of the field
theory, we are free to choose any inertial observer's rest frame
in which to evaluate the above quantities. The modes of the quantum field
will then be defined relative to this frame. We may therefore rewrite
Eq.~(\ref{eq:HATF/2D}) in the more manifestly covariant form
\begin{equation}
{\tau_0 \over \pi} \int_{-\infty}^{\infty}
{{- T_{\mu\nu} {u^\mu} {n^\nu} d{\tau}}\over {{\tau}^2+{\tau_0}^2}}
\geq -{1 \over {16 \pi {\tau_0}^2}}\,, \label{eq:CFHAT}
\end{equation}
where $\tau$ is the observer's proper time. Now consider an observer who
moves along the $-x$ direction, i.e. opposite to the direction of the
allowed modes, so that $V\rightarrow -|V|$. The flux seen
by this observer is given by
\begin{equation}
F = {\gamma}^2 \left[|V| T_{tt}-(1+V^2) T_{tx}+|V| T_{xx}\right].
                                                       \label{eq:F(-V)}
\end{equation}
Substitute Eq.~(\ref{eq:F(-V)}) into Eq.~(\ref{eq:CFHAT}), and use
Eq.~(\ref{eq:TAULAMB}) as before. Now let \hbox{$V\rightarrow -1$,}
and divide both sides by ${\gamma}^2$. Then the null limit
is again Eq.~(\ref{eq:NQI1}).

    In the present derivation we have so far assumed that all the allowed
modes were propagating in the same direction. If we now lift this restriction
by considering a general quantum state with modes propagating in both
directions, then by our earlier arguments we can see that the modes which
propagate in the same direction as our chosen null geodesic contribute
nothing to the integral. The only contribution is from modes moving
in the opposite direction. Therefore for a general quantum state
we again obtain Eq.~(\ref{eq:NQI1}).

\section{A 4D Inequality for Energy Density}
\label{sec:4DQI}
    In this section we will prove a QI on energy density
for a quantized free, minimally-coupled, massless scalar field in
four-dimensional (uncompactified) Minkowski spacetime, which is analogous to a
similar inequality on energy flux proved in \cite{F91}. From this inequality,
we will obtain AWEC and ANEC in suitable limits.

The stress-energy tensor for the scalar field is
\begin{equation}
T_{\mu \nu} = \phi_{,\mu}\phi_{,\nu}
-{1\over 2} g_{\mu \nu}\phi_{,\alpha}\phi^{,\alpha}, \label{eq:STEN}
\end{equation}
where the $g_{\mu \nu}$'s are the flat spacetime metric coefficients
written in spherical coordinates.
The wave equation
\begin{equation}
\Box \phi =0                                  \label{eq:BOXPHI}
\end{equation}
has solutions which we take to be of the form
\begin{equation}
f_{\omega l m} = \eta_{lm}\,
{g_{\omega l}(r)\over{\sqrt{2\omega}}}\,
Y_{lm}(\theta, \varphi)\, e^{-i\omega t}.           \label{eq:MODEFS}
\end{equation}
Here the $Y_{lm}(\theta, \varphi)$ are the usual spherical
harmonics, and
\begin{equation}
\eta_{l m} = e^{i{{\pi}\over 2}(l+|m|+1)}         \label{eq:PHASE}
\end{equation}
is a convenient choice of phase factor. The functions
\begin{equation}
g_{\omega l}(r)=\omega\, \sqrt{2\over R}\,\, j_{l}(\omega r), \label{eq:gDEF}
\end{equation}
where
\begin{equation}
\int_{0}^{R} r^2\,[g_{\omega l}(r)]^2\,dr=1,
\end{equation}
are normalized spherical Bessel functions. The normalization is carried
out in a large sphere of radius $R$, where we choose vanishing boundary
conditions on the sphere, i.e.,
\begin{equation}
j_{l}(\omega r)|_{r=R}=0.                    \label{eq:BC}
\end{equation}
Our boundary condition at $r=R$ implies the following condition on $\omega$:
\begin{equation}
\omega=\omega_{n l}={z_{nl}\over R},
\end{equation}
where
\begin{equation}
j_{l}(z_{n l})=0, \qquad n=1,2,\ldots        \label{eq:BZEROS}
\end{equation}
are the zeros of the spherical Bessel functions. (Later in the calculation,
we will let $R\rightarrow\infty$.)

We expand the quantized field in terms of creation and annihilation
operators as
\begin{equation}
\phi = \sum_{\omega l m}
\Bigl( a_{\omega l m}f_{\omega l m}
+ {a^\dagger}_{\omega l m}{f^*}_{\omega l m}\Bigr).  \label{eq:PHI}
\end{equation}
Here $\sum_{\omega l m}=\,\sum_{\omega=0}^{\infty}\,\sum_{l=0}^{\infty}\,
\sum_{m=-l}^{+l}$.
The normal-ordered expectation value $\langle T_{00}\rangle$ in
$t,r,\theta,\varphi$ coordinates is \cite{MODECOMMENT}
\begin{equation}
\langle T_{00}\rangle={1\over 2}\left[\langle(\phi,_{0})^2\rangle\,+
\langle(\phi,_{r})^2\rangle\, + {1\over r^2} \langle(\phi,_{\theta})^2\rangle+
\,{1\over {r^2\,{sin}^2\theta}}\langle(\phi,_{\varphi})^2\rangle\right].
                                                      \label{eq:EXP T}
\end{equation}
In flat 4D spacetime, because of translational invariance,
we are free to evaluate $\langle T_{00}\rangle$ at any point. To simplify
the calculation, we choose this point to be $r=0$. Therefore, we can use the
fact that
\begin{equation}
j_{l}(x) \simeq {x^l\over {(2l+1)!!}}\,, \qquad {\rm for}\, |x|\ll l\, ,
                                                \label{eq:JLIM}
\end{equation}
where $(2l+1)!!=1\times 3 \times 5 \times \ldots (2l+1)$, and $x=\omega r$.
Therefore, we have that
\begin{equation}
g_{\omega l, r}=\omega\, \sqrt{2\over R}\, [j_{l}(\omega r)],_{r}
\, \simeq \, {{\omega}^{l+1} \over {(2l+1)!!}}\,\sqrt{2\over R}\, l\,r^{l-1}.
                                           \label{eq:g,r}
\end{equation}

The result of evaluating $\langle T_{00}\rangle$ at $r=0$ is that only
the first two $l$-modes will contribute to our calculation. A
straightforward but tedious calculation gives
\begin{eqnarray}
\hat\rho &\equiv& {t_0 \over \pi}\, \int_{-\infty}^{\infty}\,
{{\langle T_{00}\rangle\, dt}
\over {t^2+{t_0}^2}}                    \nonumber \\
&=& {1 \over {12\pi R}}\,Re \sum_{\omega \omega'}
{\Bigl(\omega\omega'\Bigr)}^{3/2}
\Biggl\{ 3\Bigl[ \langle{a^\dagger}_{\omega' 0 0}\,a_{\omega 0 0}\rangle
e^{-|\omega -\omega'|t_0}
+ \langle a_{\omega' 0 0}\,a_{\omega 0 0}\rangle e^{-(\omega
+\omega')t_0}\Bigr]                        \nonumber \\
&+& \Bigl[ \langle{a^\dagger}_{\omega' 1 0}\,a_{\omega 1 0}\rangle\,+
\langle{a^\dagger}_{\omega' 1 1}\,a_{\omega 1 1}\rangle\,
+\langle{a^\dagger}_{\omega' 1, -1}\,a_{\omega 1, -1}\rangle\Bigr]
e^{-|\omega -\omega'|t_0}            \nonumber \\
&+& \Bigl[ \langle a_{\omega' 1 0}\,a_{\omega 1 0}\rangle\,
+\langle a_{\omega' 1, -1}\,a_{\omega 1 1}\rangle\,
+ \langle a_{\omega' 1 1}\,a_{\omega 1, -1}\rangle\Bigr] e^{-(\omega
+\omega')t_0}\Biggr\}.                     \label{eq:RHO/HAT}
\end{eqnarray}
Recall that we have standing waves inside a sphere of radius $R$,
whose size we will eventually allow to go to infinity. Therefore,
we may write the asymptotic form of the zeros of the spherical Bessel
functions $z_{n l}$, for $n$ large compared to $l$ as \cite{A-S}:
\begin{eqnarray}
z_{n l} &\sim& \Biggl(n+{l\over 2}\Biggr)\pi \sim n \pi,
\quad {\rm for}\,\,n \gg l,                              \nonumber \\
\omega_{n l} &=& {{z_{n l}} \over R} \sim {\pi n \over R}, \quad
{\rm for}\,\, \omega \gg {l\over R}\, .            \label{eq:ZRINF}
\end{eqnarray}
Let us also write the sum,
\begin{eqnarray}
&&Re \sum_{\omega \omega'}
{\Bigl(\omega\omega'\Bigr)}^{3/2}
\langle{a^\dagger}_{\omega' l m}\,a_{\omega l m}\rangle
e^{-|\omega -\omega'|t_0}               \nonumber \\
&&=Re \sum_{\omega \omega'}
\sqrt{\omega\omega'}\, |B_{\omega}B_{\omega'}|
\langle{a^\dagger}_{\omega' l m}\,a_{\omega l m}\rangle
e^{-|\omega -\omega'|t_0},                      \label{eq:RESUM}
\end{eqnarray}
where $B_{\omega}=\omega$. The right-hand side of Eq.~(\ref{eq:RESUM})
has the same form as the left-hand side of Eq. (2.26) of Ref. \cite{FR92}.
With the arguments following Eq. (2.26) of that paper and the fact that
in the large $R$ limit, $\omega \propto n$ from Eq.~(\ref{eq:ZRINF}),
one can show that
\begin{eqnarray}
Re \sum_{\omega \omega'}
{\Bigl(\omega\omega'\Bigr)}^{3/2}
\langle{a^\dagger}_{\omega' l m}\,a_{\omega l m}\rangle
e^{-|\omega -\omega'|t_0}               \nonumber \\
\geq \sum_{\omega \omega'}
{\Bigl(\omega\omega'\Bigr)}^{3/2}
\langle{a^\dagger}_{\omega' l m}\,a_{\omega l m}\rangle
e^{-(\omega +\omega') t_0}.                       \label{eq:RESUMINQ}
\end{eqnarray}

{}From Eqs.~(\ref{eq:RHO/HAT}) and ~(\ref{eq:RESUMINQ}), we may write
\begin{eqnarray}
\hat\rho &\equiv& {t_0 \over \pi}\, \int_{-\infty}^{\infty}\,
{{\langle T_{00}\rangle\, dt}
\over {t^2+{t_0}^2}}                     \nonumber \\
&\geq& \,{1 \over {12\pi R}}\,Re \sum_{\omega \omega'}
{\Bigl(\omega\omega'\Bigr)}^{3/2}
\Bigl[3(\langle{a^\dagger}_{\omega' 0 0}\,a_{\omega 0 0}\rangle
+ \langle a_{\omega' 0 0}\,a_{\omega 0 0}\rangle)         \nonumber \\
&+& \langle{a^\dagger}_{\omega' 1 0}\,a_{\omega 1 0}\rangle\,+
\langle a_{\omega' 1 0}\,a_{\omega 1 0}\rangle\,+
\langle{a^\dagger}_{\omega' 1 1}\,a_{\omega 1 1}\rangle\,+
\langle a_{\omega' 1 1}\,a_{\omega 1, -1}\rangle\,      \nonumber \\
&+& \langle{a^\dagger}_{\omega' 1, -1}\,a_{\omega 1, -1}\rangle
\, + \langle a_{\omega' 1, -1}\,a_{\omega 1 1}\rangle\Bigr]\,
e^{-(\omega + \omega')t_0}                     \nonumber \\
&=& {1 \over {12\pi R}}\,Re\sum_{{\omega \omega'}\atop{l l', m m'}}\,
h_{\omega l}\,h_{\omega' l'}\,
(\langle a^\dagger_{\omega' l' m'}\,a_{\omega l m}\rangle
+ \langle a_{\omega' l, -m}\,a_{\omega l m}\rangle)\,,   \label{eq:RHOHATA'S}
\end{eqnarray}
where
\begin{equation}
h_{\omega l}=\left\{ \begin{array}{lll}
           \sqrt{3}\,\,{\omega}^{3/2}\,\,e^{-\omega t_0}, & \mbox{for $l=0$} \\
           {\omega}^{3/2}\,\,e^{-\omega t_0}, & \mbox{for $l=1$} \\
           0\,,                             & \mbox{for $l>1$}.
                         \end{array}                 \label{eq:HARRAY}
                  \right.
\end{equation}
Now use Eqs. (2.34) and (2.42) of Ref. \cite{FR92} to write
\begin{equation}
\hat\rho \geq
-{1 \over {24\pi R}}\,\sum_{\omega l}\, (2l+1)\,{h^2_{\omega l}}.
                                                   \label{eq:RHOHATH'S}
\end{equation}
If we perform the sum over $l$, we have
\begin{equation}
\sum_{\omega l}\, (2l+1)\,{h^2_{\omega l}}=h^2_{\omega 0}\,+3h^2_{\omega 1}
=6\, {\omega}^{3}\,e^{-2 \omega t_0},                 \label{eq:HSUM}
\end{equation}
and therefore
\begin{equation}
\hat\rho \geq -{1 \over {24\pi R}}\,\sum_{\omega}
6\, {\omega}^{3}\,e^{-2 \omega t_0}.             \label{eq:RHOHAT-OSUM}
\end{equation}
Now let us use the fact that $\omega \sim {(\pi n/R)}$ for $R$ large, to
write\linebreak
 $\sum_{\omega}\rightarrow (R/ \pi)\, \int_{0}^{\infty}\, d{\omega}$,
as $R\rightarrow \infty$. Thus
\begin{equation}
\hat\rho \geq -{1 \over {4{\pi}^2}}\,\int_{0}^{\infty}\, d{\omega}\,
{\omega}^{3}\,e^{-2 \omega t_0}\,.           \label{eq:RHOHAT-WINT}
\end{equation}
An evaluation of the integral gives us our desired result
\begin{equation}
\hat\rho \geq -{3\over {32 {\pi}^2 {t_0}^4}}\,,  \label{eq:HATRHO/QI}
\end{equation}
for all $t_0$. This inequality has the same form as a similar
inequality for the energy flux seen by a timelike inertial observer in
4D flat spacetime, Eq.~(\ref{eq:HATF/4D}), which was derived
in Ref. \cite{F91}.

Although we derived our bound for the case where the observer's velocity,
$V$, was equal to zero, in fact our result is more general. To see this,
consider the following argument.
We chose to do the calculation in the frame of reference of an observer
at rest at $r=0$. However, from the underlying Lorentz-invariance of the
field theory, we could have chosen any inertial frame as ``the rest frame''
and done the calculation in that frame. The mode functions used in the
derivation of Eq.~(\ref{eq:HATRHO/QI}) would then simply be defined
relative to whatever inertial frame we choose. In this chosen frame
$\langle T_{\mu\nu} u^{\mu} u^{\nu}\rangle=\langle T_{00}\rangle$.
Since the bound we derived holds in {\it any} such inertial frame, we
may write our QI for 4D uncompactified spacetime, Eq.~(\ref{eq:HATRHO/QI}),
in the more manifestly covariant form:
\begin{equation}
\hat \rho = {{\tau_0} \over \pi}\, \int_{-\infty}^{\infty}\,
{{\langle T_{\mu\nu} u^{\mu} u^{\nu}\rangle\, d\tau}
\over {{\tau}^2+{\tau_0}^2}} \geq
-{3\over {32 {\pi}^2 {\tau_0}^4}}\,,  \label{eq:4DENQI}
\end{equation}
for all $\tau_0$, where $u^{\mu}$ is the tangent vector to the timelike
geodesic (i.e., the observer's four-velocity) and $\tau$ is the observer's
proper time.

In the limit $\tau_0\rightarrow \infty$, corresponding to the width of our
sampling function going to infinity, we sample the entire geodesic and
obtain
\begin{equation}
\int_{-\infty}^{\infty}\,
\langle T_{\mu\nu} u^{\mu} u^{\nu}\rangle\, d\tau \geq\,0.  \label{eq:4DAWEC}
\end{equation}
This relation holds for all timelike geodesics.
As in two-dimensions, we again find that we can derive AWEC from the QI
for timelike geodesics.

   Recall that in 2D uncompactified Minkowski spacetime, we were able to
derive a QI for null geodesics, Eq.~(\ref{eq:NQI1}). It is not clear that an
analogous relation exists in four dimensions. One might naively expect
it to take the form of Eq.~(\ref{eq:NQI1}), but with the right-hand-side
proportional to $\lambda_0^{-4}$. However, such a relation would not
be invariant under rescaling of the affine parameter and hence does not
seem to be meaningful. An attempt to derive a QI in four dimensions
starting with null geodesics {\it ab initio} cannot employ the techniques
we have used to obtain
Eq.~(\ref{eq:4DENQI}), because the latter is based upon a mode expansion
in the observer's rest frame.
Nonetheless, it is possible to derive ANEC in four-dimensional Minkowski
spacetime,
\begin{equation}
\int_{-\infty}^{\infty}\,
\langle T_{\mu\nu} K^{\mu} K^{\nu}\rangle\, d\lambda \geq\,0\, ,
                                                   \label{eq:ANEC4D}
\end{equation}
as the null limit of Eq.~(\ref{eq:4DAWEC}).  One uses a procedure
\cite{NULLLIMITCOMMENT} analogous
to that outlined after Eq.~(\ref{eq:TAULAMB}), which was used to derive
Eq.~(\ref{eq:DHATTKK}) from Eq.~(\ref{eq:DHATRIGHT}).

\section{Conclusions}
\label{sec:summary}
   In this paper, we have uncovered deep connections between QI-type
restrictions on negative energies and averaged energy conditions.
As in the case of the averaged energy conditions, the QI bounds in
this paper are all formulated covariantly.
In a 2D spatially compactified Minkowski spacetime with circumference $L$,
for a quantized massless scalar field, we defined a difference
of the expectation values of $T_{\mu\nu} u^{\mu} u^{\nu}$ in an arbitrary
quantum state and in the Casimir vacuum state. (Here $u^{\mu}$ is the
two-velocity of an arbitrary inertial observer.) It was then shown that this
difference satisfied a QI-type bound. From this bound, it was shown that
the difference in expectation values also satisfies an AWEC-type integral
condition. In the $L\rightarrow\infty$ limit, we obtained both a QI bound
and AWEC for 2D
uncompactified flat spacetime. Similar QI's for null geodesics, in
both 2D compactified and ordinary 2D Minkowski spacetime,
were derived. Again, it was found that the
difference in expectation values satisfied a QI-type bound for null
geodesics, which in an appropriate limit reduces to ANEC.
In the $L\rightarrow\infty$ limit, we obtained a QI bound and ANEC for 2D
uncompactified spacetime. These results are surprising since it is
known that for
$\langle T_{\mu\nu} u^{\mu} u^{\nu} \rangle$ in the renormalized
Casimir vacuum state alone, the timelike and null QI-type bounds, as well
as AWEC and ANEC, are {\it not} satisfied.

How should one physically interpret this difference inequality? In
the case of 2D compactified Minkowski spacetime, one has two ways of
enhancing the effects of negative energy. These consist of: a) changing
the size of the space $L$, thereby altering the background Casimir vacuum
state, and b) changing the quantum state, $|\psi\rangle$, of the field.
Our difference inequality seems to imply that b) is not very effective
at magnifying the effects of negative energy over and above that of the
negative Casimir vacuum background energy. If one shrinks the size of the
space (figuratively speaking, of course), the energy density grows more
negative but over a region of smaller size (i.e., a universe with a
smaller value of $L$). This would again seem to restrict the production
of large-scale effects through the manipulation of negative energy. It
would be interesting to see if the behavior of our simple 2D model is
suggestive of the general case. It may be possible to construct similar
``difference inequalities'' in 4D curved spacetimes where ANEC and AWEC
are violated. The curvature would be expected to couple to the
expectation value of $T_{\mu\nu}$ rather than to a difference of
expectation values. Nevertheless, if such difference inequalities
exist in these cases,
they may perhaps place limits on the degree of averaged energy condition
violation.

A covariant QI-type bound on the energy density was also derived for
a quantized, minimally-coupled, free massless scalar field
in 4D Minkowski spacetime. It implies that an inertial observer in 4D
flat spacetime cannot see unboundedly large negative energy densities for
an arbitrarily long period of time. Although our result was proved
for free massless
fields in flat spacetime, it suggests that the manipulation of negative energy
to build a ``warp-drive'', at least using the procedure
suggested in Ref. \cite{WARP}, may be extremely difficult if not impossible.
If one starts in flat space and attempts to collect enough
negative energy to create the necessary spacewarp, our QI suggests that
compensating $(+)$ energy will arrive before enough $(-)$ energy is
collected to significantly curve the space. Perhaps this restriction
could be circumvented by starting, for example, with other kinds of
(not necessarily free or massless) fields in curved spacetime.

Is it possible that within the realm of
semiclassical gravity theory, the presence of negative energy densities
due to quantum effects might invalidate the singularity theorems, before
quantum gravity effects become important? Although this seems unlikely,
there are presently no firm proofs one way or the other.
It remains possible that although ANEC
might fail in some regions of a given spacetime (e.g., along half-infinite
null geodesics in an evaporating
black hole spacetime), it may hold in enough
other regions that, for example, the conclusions of Penrose's singularity
theorem might still be valid \cite{TR,B/TC,FR/IN PREP}.
In regions of evaporating black hole spacetimes where ANEC is violated,
it may be possible to get a (more limited) QI-type bound that
measures the degree of ANEC violation and which
is also scale-invariant. The existence of such an inequality
may depend on the presence of a characteristic length scale,
e.g. the mass of the black hole \cite{FR/IN PREP}.

It should also be noted
that recent work of Kuo and Ford \cite{KF,Kuo} indicates that in
flat spacetime,
negative energy densities are subject to large fluctuations. This suggests
that the naive use of the semiclassical theory of gravity may be suspect,
at least in some situations involving negative energy.

\vskip 0.2 in
\centerline{\bf Acknowledgements}
We would like to thank Arvind Borde and Paul Davies for useful discussions.
TAR would like to thank the members of the Tufts Institute of Cosmology
for their kind hospitality and encouragement while this work was being done.
This research was supported in part by NSF Grant No. PHY-9208805 (to LHF).


\begin{thebibliography}{--}
\bibitem{CSCALE} Note that here we are discussing multiplicative rescaling
of the affine
parameter. Although we have set $c_0=0$, this merely fixes the origin of the
affine parameter. A non-zero choice of $c_0$ leaves $\lambda_0$ and
$\lambda_p$ unchanged, and simply translates the position of the peak of the
sampling function. Hence the content of Eq.~(\ref{eq:DHATTKK2}) is unchanged.
\bibitem{NULLLIMITCOMMENT} This ``timelike to null'' limit works in
            Minkowski spacetime, which has a simple causal structure.
            Such a limit need not be valid in spacetimes with a more
            complicated causal structure.
\bibitem{MODECOMMENT} The reason for the technical complication of working
        with spherical modes is that, in the case of the energy density in
        4D, we were not able to find a simple analog to the ``rotation of
        modes'' argument which was used in \cite{F91} to derive
        the 4D flux inequality.
\end{thebibliography}
\end{document}